\documentclass[pdf,aoas,preprint]{imsart}

\RequirePackage[OT1]{fontenc}
\RequirePackage{amsthm,amsmath}
\RequirePackage[numbers]{natbib}
\RequirePackage[colorlinks,citecolor=blue,urlcolor=blue]{hyperref}
\RequirePackage[T1]{fontenc}
\RequirePackage{color}
\RequirePackage{placeins}
\RequirePackage{comment}
\RequirePackage{layouts}
\RequirePackage{graphicx}
\RequirePackage{float}

\startlocaldefs
\numberwithin{equation}{section}
\theoremstyle{plain}

\endlocaldefs
\newcommand*\cpp{C\kern-0.2ex\raisebox{0.4ex}{\scalebox{0.8}{+\kern-0.4ex+}}}

\DeclareMathOperator*{\med}{med}
\DeclareMathOperator*{\mad}{mad}
\DeclareMathOperator*{\cov}{cov}
\DeclareMathOperator*{\ave}{ave}

\expandafter\def\expandafter\normalsize\expandafter{%
    \normalsize
\setlength\abovecaptionskip{.25cm}
\setlength\belowcaptionskip{-.25cm}
\setlength\abovedisplayskip{.25cm}
\setlength\belowdisplayskip{.25cm}
\setlength\belowdisplayshortskip{.25cm}
\setlength\abovedisplayshortskip{.25cm}
}

\definecolor{fmcd}{RGB}{255,0,0}
\definecolor{fmve}{RGB}{0,0,255}
\definecolor{msde}{RGB}{0,255,0}
\definecolor{PCS}{RGB}{255,0,255}

\begin{document}

\begin{frontmatter}
\title{Finding Multivariate Outliers With FastPCS}
\runtitle{The PCS Index of Multivariate Outlyingness}

\begin{aug}
\author{\fnms{Kaveh} \snm{Vakili}\ead[label=e1]{kaveh.vakili@wis.kuleuven.be}} \and
\author{\fnms{Eric} \snm{Schmitt}\ead[label=e2]{eric.schmitt@wis.kuleuven.be}}

\runauthor{K. Vakili and E. Schmitt}


\address{
\printead{e1}\\
\phantom{E-mail:\ }\printead*{e2}\\
}


\end{aug}

\begin{abstract}
The Projection Congruent Subset (PCS)  
is a new method for finding multivariate outliers. 
Like many other outlier detection procedures, PCS 
searches for a subset which minimizes a criterion. 
The difference is that the new criterion was designed
 to be insensitive to the outliers. PCS is supported 
by FastPCS, a fast and affine equivariant algorithm 
which is also detailed. Both an extensive simulation
 study and a real data application from the field of
 engineering show that FastPCS performs better than 
its competitors.\end{abstract}



\end{frontmatter}

\section{Introduction}

Outliers are observations that
 do not follow the pattern
 of the majority of the data \citep{mcs:RZ90}.
Outlier identification is a major part 
 of data analysis for at least two reasons. 
First, because a few outliers, if left unchecked, will exert a disproportionate pull 
on estimated parameters, and we generally  
do not want our inferences to depend
 on such observations. In addition, we may want to find outliers 
 to set them aside and study them as objects of interest in their own right. 
In any case, detecting outliers in settings involving more than two variables 
is difficult because we can not inspect the data visually and have to rely on 
algorithms instead.

Formally, this paper concerns 
itself with the simplest, 
most basic variant of the 
multivariate outlier 
detection problem. The general
 setting is that of a sample of 
$n$ observations $x_i\in\mathbf{R}^p$ 
(with $n>p$), at least 
$h=\lfloor(n+p+1)/2\rfloor$ of
 which are drawn from a multivariate 
elliptical distribution. 
 The objective is to identify
 reliably the index
 of the remaining ones. 
A more complete treatment of this
 topic can be found 
in textbooks \citep[for example]{mcs:MMY06}.

In this article we introduce
 PCS, a new procedure for 
finding multivariate outliers.
 We also detail FastPCS, a fast 
 algorithm for computing it. 
The main output of FastPCS is an outlyingness 
index measuring how much each observation
 departs from the pattern set by the
 majority of the data.

The PCS outlyingness index is affine
 equivariant (meaning that the 
 outlyingness ranking of the 
observations is not affected by
 a linear transformation of the
 data) and can be computed 
efficiently for moderate values 
of $p$ and large values of $n$.

 To derive this index, FastPCS 
 proceeds in two steps. 
First, it strives to select 
among many possible $h$-subsets  
of observations one devoid of
 outliers. Then, the outlyingness
 index is simply the distance of
 each observation to this subset.

For easier outlier detection 
problems, we find that 
our approach produces results similar 
to state-of-the-art  
outlier detection algorithms. 
When considering more difficult
 cases however we find that the
 solution we propose leads to 
significantly better outcomes. 

In the next section we
 motivate and define the PCS outlyingness
 and FastPCS. 
Then, in Section 3 we compare FastPCS 
to several competitors on 
synthetic data. In Section 4
 we conduct a real data comparison.
 In Section 5, we offer closing
 thoughts and discuss directions
 for further research. 

\section{The PCS outlyingness index}
\subsection{Motivation}
Throughout this note, we will
 denote as $H_m$ a subset of
 the indexes $\{1,2,\ldots,n\}$ 
(The subscript $m$ indexing such sets will be used later on). 
For any $H_m$ we denote its (sample) mean and
 covariance matrix as 
\begin{equation}\label{mcs:raw}
(t_m,S_m)=\left(\ave_{i\in H_m}x_i,\cov_{i\in H_m}x_i\right)
\end{equation} 
and we will write the squared Mahalanobis 
distance of an observation $x_i$ as 
\begin{equation}
   d_{MD,i}^2(t_m,S_m) = (x_i-t_m)'S_m^{-1}(x_i-t_m)\;.
\end{equation}  

FastPCS starts selects among 
many such subsets an $h$-subset of
 observations that minimizes a 
criterion. Other affine equivariant
 outlier detection algorithms that 
proceed in this way are the 
FastMVE \citep{mcs:MMY06}, 
the FastMCD \citep{mcs:RV99} 
 and the SDE \citep{mcs:S81}. 
For all four procedures we will denote 
the chosen $h$-subset as $H_*$. 

Each of these algorithms
 relies on a different criterion to select 
the optimal $h$-subset. In all cases, there is a
 straightforward relationship between
 $H_*$ and the resulting outlyingness
 index. For example, for FastMVE and 
FastMCD the outlyingness index is the
 vector of distances $d_{MD,i}^2(t_*,S_*)$. 
For SDE the outlyingness index is the
 vector of values of $\text{P}_M(x_i)$ 
\begin{eqnarray}\label{mcs:sde1}
\text{P}_M(x_i)=\displaystyle\sup_{a_m\in B_{M_p}}\;\;\;\frac{|x_i'a_m-\med_{j=1}^{n}(x_j'a_m)|}{\mad_{j=1}^{n}(x_j'a_m)}\;,
\end{eqnarray}
where $B_{M_p}$ is a set of $M_p$ 
directions orthogonal to  
hyperplanes spanned by a 
$p$-subset of $\{1:n\}$ 
 and $H_*$ contains the
 observations with smallest values of 
$\text{P}_M(x_i)$. 

In all cases, if $H_*$ itself 
is contaminated, then the resulting
 outlyingness index can no longer
  be depended upon to find the 
 outliers. For FastMVE and FastMCD, 
$H_*$ is the subset of $h$ 
observations with smallest 
covariance matrix determinant and enclosed by the ellipsoid
 with smallest volume, respectively. Given enough outliers,
 an adversary can easily confound these two 
characterizations (for example, by placing
 outliers on a subspace of $\mathbf{R}^p$) and ensure
 that a contaminated $h$-subset will
 always be chosen over $h$-subsets formed of 
genuine observations. 

\begin{figure}[h!]
\centering
\includegraphics[width=1\textwidth]{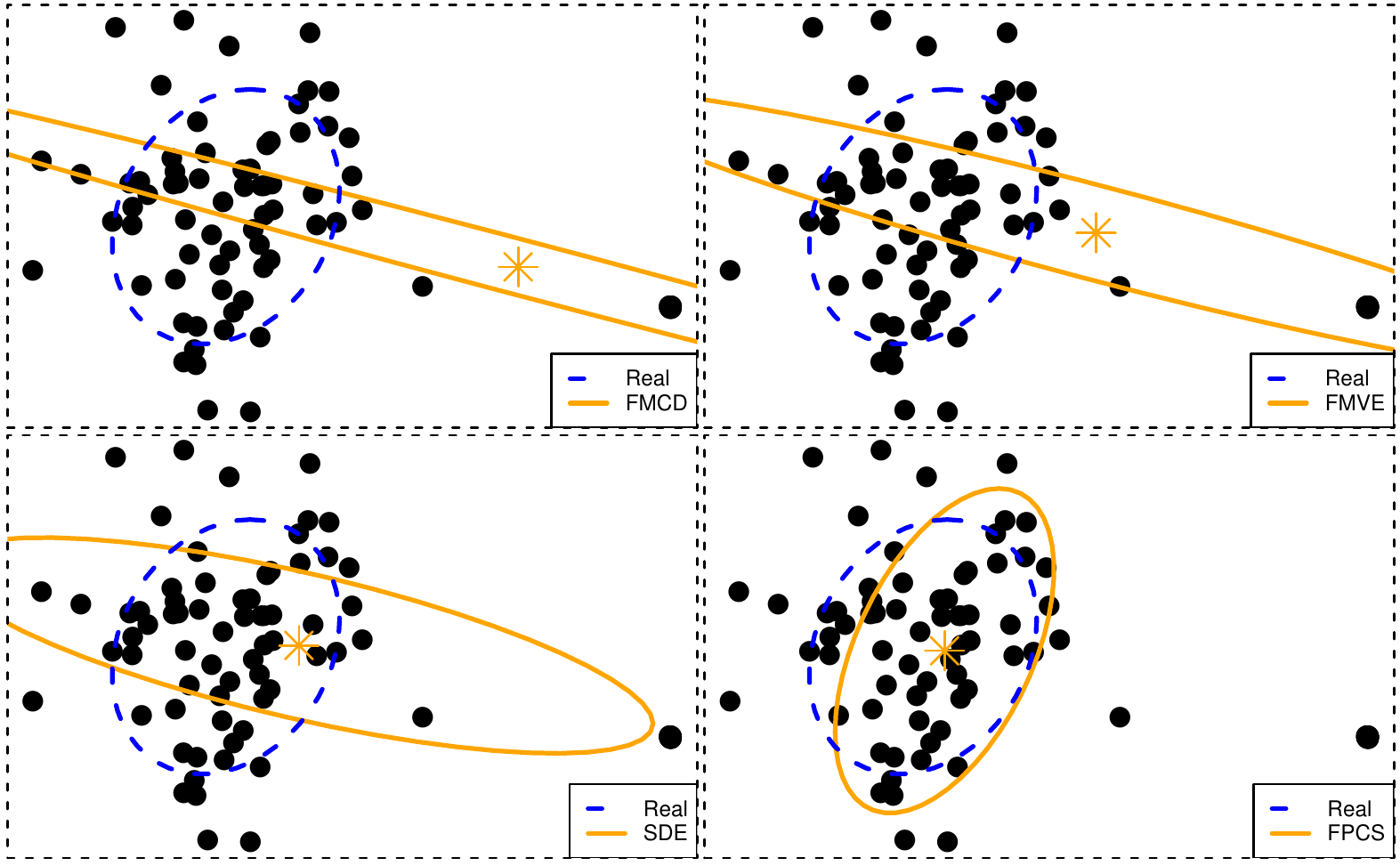}
	\caption{The four plots depict (left right, top, bottom) the 
same configuration of observations. The darker blue dashed ellipses 
show the contours of the model governing the distribution of the 
majority --here 70 out of 100-- of the observations. The light orange, solid, ellipses 
 show, respectively,  the FastMCD, FastMVE, SDE and FastPCS estimates of scatter and location.} 	
\label{mcs:f1a}
\end{figure}

It is not always recognized that the
 SDE outlyingness index is also 
sensitive to the outliers. Given enough outliers,
 they can always be placed in 
such a way that the denominator in 
Equation $\eqref{mcs:sde1}$ will be
 smaller for some direction $a_*$ along
 which the numerator is smaller for 
outliers than for genuine observations
 (for example, consider a direction $a_*$
 close to the principal axis of the lighter, orange
 ellipse in the lower-left panel in Figure~\ref{mcs:f1a}).
 Then, for many observations the maximum 
outlyingness will be attained at $a_m=a_*$.
 Repeated over a large number of such 
directions, this will cause the value of
 the outlyingness index in Equation 
$\eqref{mcs:sde1}$ to be smaller for many
 outliers than for genuine observations.

Consider the following example. The four 
panels in Figure~\ref{mcs:f1a} all depict
 the same 70 draws from a standard bivariate
 normal distribution together with 30 draws
 from a second normal distribution with 
smaller variance and centered at $(5,-1)$.
 Light, orange ellipses in the first three plots 
show the $(t_*,S_*)$ obtained using the 
best subset found by the FastMCD, FastMVE
 and SDE. These were computed with the R 
package \texttt{rrcov} \citep{mcs:TF09}, using default settings; 500 starting $p$-subsets 
(for the first two), and 500 directions $a_m$
 (for SDE); and $h=51$ (see Section~\ref{mcs:s5}) in all cases.

In all three cases, the fitted ellipses 
(drawn in solid orange lines), fail to 
adequately fit any $h$-subset of the data,
 including the subset they enclose. In 
particular, their centers of symmetry 
(orange stars) are not located in areas of
 highly concentrated observations. In all 
cases, the model fitted on $H_*$ appears
 visually distinct from the distribution
 governing the good part of the data 
(drawn as a dashed, dark blue ellipse). This is
 confirmed by the biases (a dimensionless 
 measure of dissimilarity between two subsets,
 see Section~\ref{mcs:s3}) of 7, 5 and 3.5 for
 the three algorithms.

The algorithm we propose differs
 from these methods in that it 
uses a new measure of multivariate 
congruence (which we detail in the next 
section) to select the optimal $h$-subset.   
The main advantage of this new characterization
 lies in its insensitivity to the outliers.
 As we argue below, this makes the
 outlyingness index derived from FastPCS both 
quantitatively and qualitatively
 more reliable.

\subsection{Construction of the PCS Outlyingness index}

PCS looks for the $h$-subset
 of multivariate observations
 that is most {\em congruent} 
along many univariate
projections. In this
context, we measure the 
congruence of a given
 $h$-subset along a given
 projection by the size of 
its overlap with a second
 subset that is optimal 
(in a sense we make precise below) 
on that projection.
 The PCS 
criterion is based on the observation
 that a spatially cohesive 
$h$-subset will tend to be
 congruent with these optimal 
$h$-subsets, and a spatially
 disjoint one will not.

More precisely, denoting by $A_{mk}$ 
a $p \times p$ matrix formed of $p$ observations 
(we detail below how we pick these $p$ observations),
 $a_{mk}$ the hyperplane $\{a_{mk}:A_{mk}a_{mk}=1_p\}$
and $d_{P,i}$ the (squared) orthogonal distance of 
 $x_i$ to $a_{mk}$:
\begin{equation}\label{mcs:crit0}
    d_{P,i}^2(a_{mk}) = (x_i'a_{mk}-1)^2/||a_{mk}||^2\;.
\end{equation}
 The set of $h$ observations with
 smallest $d_{P,i}^2(a_{mk})$ will 
be denoted as $H_{mk}$. For a given
 subset $H_m$ and direction $a_{mk}$ we
define the {\em incongruence index} 
of $H_m$ along $a_{mk}$ as
\begin{equation}\label{mcs:crit1}
   I(H_m,a_{mk}):=
    \log\displaystyle\ave_{i\in H_m}d^2_{P,i}(a_{mk})-\log\displaystyle\ave_{i\in H_{mk}}d_{P,i}^2(a_{mk})\;.
\end{equation}
This index is always positive
 and will have small value if 
 the projection of the members 
of $H_m$ along $a_{mk}$ greatly
 overlaps with the members of $H_{mk}$. 
To remove the dependence of Equation
 \eqref{mcs:crit1} on $a_{mk}$ we 
measure the incongruence of $H_m$ by considering
 the average over many directions:
\begin{equation}\label{mcs:crit2}
      I(H_m):=\ave_{a_{mk}\in B(H_m)} I(H_m,a_{mk})\;,
\end{equation}
where $B(H_m)$ are all directions 
orthogonal to a hyperplane spanned
by a $p$-subset of $H_m$. We call 
the $H_m$ with smallest $I(H_m)$ 
the {\em projection congruent subset}.
In essence, the $I$ index measures
 the spatial cohesion of an $h$-subset $H_m$ in 
terms of how much the projections
 of its members overlap with those
 of the $H_{mk}$ over many projections.
In practice, it would be  too laborious 
 to evaluate Equation \eqref{mcs:crit2}
 over all members of $B(H_m)$. 
A practical solution is to take 
 the average over a sample of $K$
 random directions $\tilde{B}_K(H_m)$ 
instead.

The $I$ index of a spatially disjoint
$h$-subset tends to be higher than 
that of a spatially cohesive $h$-subset.
 This is because when $H_m$ forms a 
spatially cohesive set of observations,
 $\#\{H_m\cup H_{mk}\}$ tends
 to be larger.

\begin{figure}[h!]
\centering
\includegraphics[width=0.49\textwidth]{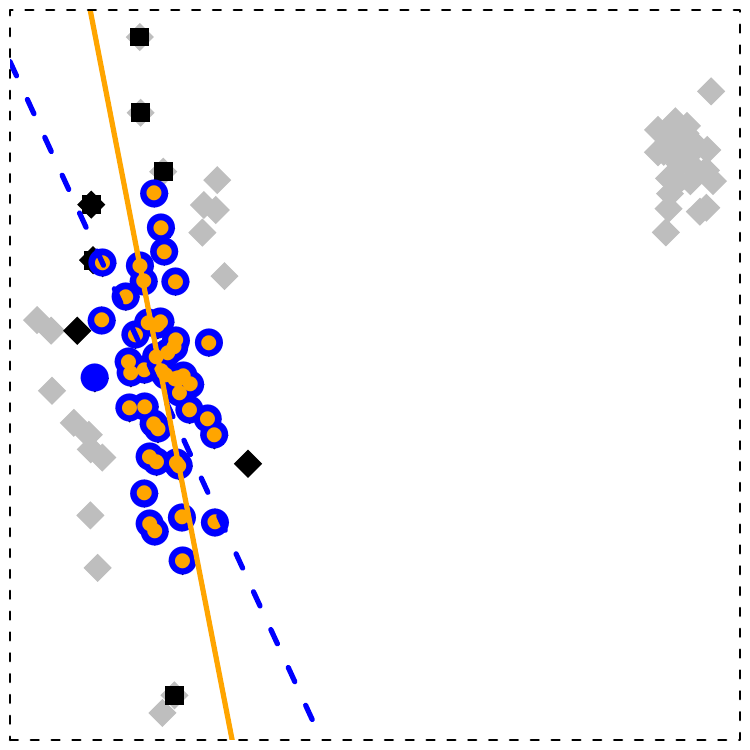}
\includegraphics[width=0.49\textwidth]{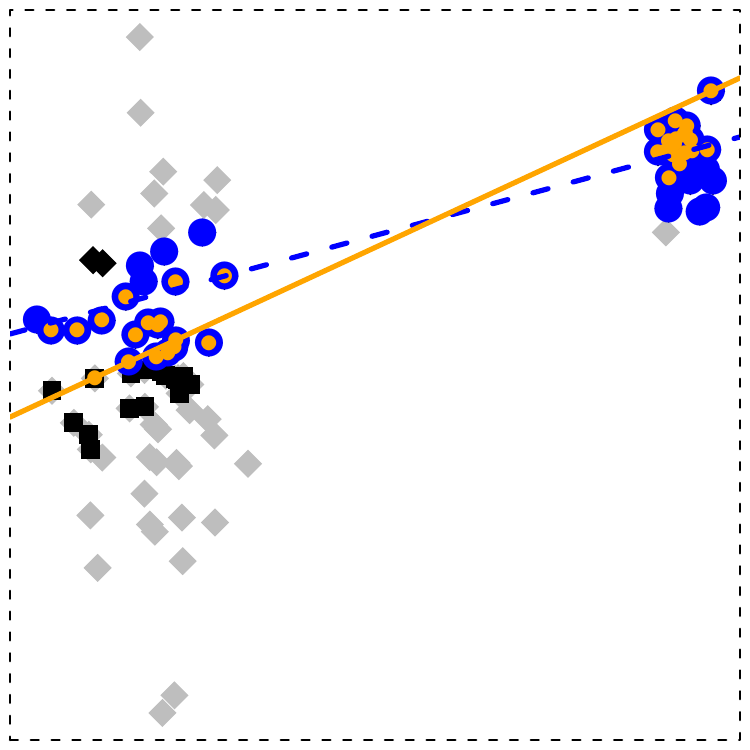}
	\caption{Incongruence index for a subset $H_1$ of spatially cohesive observations (left) and 
a subset $H_2$ of spatially disjoint observation (right)} 	
\label{mcs:f1b}

\end{figure}

 This is illustrated on an example in 
Figure \ref{mcs:f1b}. Both panels depict
 the same set of $n=100$ points. These 
points form two disjoint groups of 
observations. The main group contains 
70 points and is located on the left-hand side. Each panel illustrates the behavior
 of the $I$ index for a given $h$-subset of
 observations. $H_1$ (left) forms a set of 
spatially cohesive observations, all belonging
 to the main group. $H_2$, in contrast, forms
 a spatially disjoint set of observations and
 contains members drawn from both groups. 

For each $H_m$-subset, $m=\{1,2\}$, we drew two 
hyperplanes $a_{m1}$ (dark blue, dashed) and $a_{m2}$
 (light orange). 
The dark blue dots show the members of $\{H_m\cup H_{m1}\}$. Similarly, light orange dots show the 
members of $\{H_m\cup H_{m2}\}$.  The  diamonds (black squares) show the members of $H_{m1}$
 ($H_{m2}$) that do not belong to $H_m$. After 
just two projections, the number of non-overlapping 
observations (i.e.\{ $\{H_m\setminus H_{m1}\}\cap \{H_m\setminus H_{m2}\}\}$)
 is 8 ($m=1$) and 18 ($m=2$) respectively. 
As we increase the number of directions $a_{mk}$, this 
pattern repeats and the difference between a spatially 
cohesive and a disjoint subset grows steadily.

The $I$ index measures the size of this overlap.
 For a direction $a_{mk}$, the members of $H_{m1}$ and $H_{m2}$
not in $H_m$ (shown as diamonds and black squares
 in Figure \ref{mcs:f1b}) will decrease the 
denominator in Equation $\eqref{mcs:crit1}$ without
 affecting the numerator, increasing the overall 
ratio. Consequently, $h$-subsets formed of spatially disjoint 
groups of observations will have larger values of
 the $I$ index. Crucially, the 
$I$ index characterizes a cohesive $h$-subset
 of observations independently of the spatial 
configuration of the outliers. For example, the 
pattern shown in Figure \ref{mcs:f1b} would still
 hold if the cluster of outliers were more 
 concentrated. This is also illustrated in the fourth
 quadrant of Figure \ref{mcs:f1a} where the optimal 
$h$-subset found by FastPCS is not unduly attracted
 by members of the smaller cluster of 
 observations located on the right. 

In Sections \ref{mcs:s3} and \ref{mcs:s4}, we 
show that this new characterization allows FastPCS
 to reliably select uncontaminated $h$-subsets. 
This includes many situations where 
other algorithms fail to do so. 
 First though, the following section details the 
FastPCS algorithm.

\subsection{A Fast Algorithm for the PCS Outlyingness}

To compute the PCS outlyingness index, 
we propose the FastPCS algorithm 
$\eqref{mcs:fmcs}$. It combines ideas
 from FastMCD (the use of
 random $p$-subsets as starting points 
and an iterative concentration step) with some new ones. 
Throughout, $M_p$ denotes the number of starting 
$p$-subsets.  

An important characteristic of our algorithm is that it 
can detect exact-fit situations:
when $h$ or more observations lie exactly on a subspace, FastPCS
will return the indexes of an $h$-subset of those observations   
and the fit given by the observations in $H_*$ will 
coincide with the subspace. Since FastPCS is 
affine equivariant, this behavior suggests 
that it has maximal finite sample breakdown point, as discussed in \citep[remark 1, pg 123]{mcs:RL87} and \citep{mcs:R94,mcs:MMY06}.
Intuitively, this is because in any affine equivariant metric 
the $n-h$ observations not in the subspace do in effect
 lie arbitrarily far away from the ones that are. 

Step $b$ increases the size of $H_m$ from $p+1$ to its
 final size $h=\lfloor(n+p+1)/2\rfloor$ in $L$ steps, 
rather than in one as is done in FastMCD. This improves
 the robustness of the algorithm when outliers are close
 to the good data. We find that increasing $L$ does not 
improve performance much if $L$ is greater than 3 
and use $L=3$ as default.

\vskip0.15cm
\hrule
\vskip0.1cm
\noindent \begin{equation}\label{mcs:fmcs}
\text{Algorithm FastPCS}
\end{equation}
\hrule
\begin{tabbing}
\=for \= $m=1$ to $M_p$ do:\\ 
$a$:\>\> 
\begin{math}
  H_m^0\gets\{p+1 \text{ observations drawn at random from $1:n$}\}	
\end{math} \\
   $b$:\>\>for \=$l=1$ to $L$ do:\\
\>\>\>
\begin{math}
   D_i(H_m^l) \gets \displaystyle\ave_{k=1}^{K}  
    \frac{d_{P,i}^2(a_{mk})}
         {\displaystyle\ave_{j\in H_{m}^l}d_{P,j}^2(a_{mk})}
    \,\,\,\, \text{ for all } i=1,\ldots,n
\end{math}\\
\>\>\>set 
\begin{math}
  q \gets \lfloor(n-p-1)l/(2L)\rfloor+p+1			
\end{math}\\
\>\>\>set 
\begin{math}
  H_m^l \gets \left\{i: D_i(H_m^l) \le D_{(q)}(H_m^l)\right\}
      \,\,\,\, \text{ (`concentration step')} 
\end{math}\\
\>\>end for\\
\>\>
\begin{math}
H_m \gets H_m^L
\end{math}\\
   $c$:\>\>  compute
\begin{math}
  I(H_m) \gets \displaystyle\ave_{k=1}^{K} I(H_m,a_{mk})
\end{math}
\\
\>end for\\
Keep $H_*$, the subset $H_m$ with lowest $I(H_m)$. The final outlyingness index\\
 of each observation is given by $d_{MD,i}(t_*,S_*)$.
\end{tabbing}
\vskip-0.1cm
\hrule
\vskip0.15cm

Empirically also, we found that small values for $K$ 
are sufficient to achieve good results and that 
we do not gain much by increasing $K$ above 25,
 so we set $K=25$ as the default (this is 
the value we use throughout). 
That such a small number of random directions 
suffice to reliably identify the outliers is 
remarkable. This is because FastPCS  
 uses directions generated by
 $p$-subsets of $H_m$. 

Compare this to SDE algorithm 
which needs a much larger number 
of projections to reliably find
 the outliers. This is because in SDE the 
data is projected along directions
 given by hyperplanes passing through $p$ 
points drawn indiscriminately from 
the entire set of observations. 
Consequently, when the contamination
 rate is high, most of these $p$-subsets
 will be contaminated, yielding 
directions that can end up almost 
parallel to each other. In contrast, 
our choice always ensures a wider spread of 
directions when $H_m$ is uncontaminated and
 thus yields better results.

 Like FastMCD and FastMVE,
 FastPCS uses many random $p$-subsets as 
starting points. The number of initial 
$p$-subsets, $M_p$, must be large enough to 
ensure that at least one of them is 
uncontaminated. 
For FastMCD and FastMVE, for each starting $p$-subset, the computational 
complexity scales as $O(p^3+np^2)$  whereas for FastPCS it is 
$O(p^3+np)$.
 Computing the SDE outlyingness index  also costs $O(p^3+np)$ 
for each hyperplane and here also the number of such hyperplanes must
 be of order $M_p$ to ensure that at least one of them
 does not pass through any of the outliers. 

The value of $M_p$ (and therefore the computational complexity
 of all four algorithms) grows exponentially with $p$. The 
actual run times will depend on implementation choices 
but in our experience FastMCD is slightly faster than
 FastPCS and both are noticeably slower than either FastMVE or SDE.
In practice this means all four procedures become
 impractical for values of $p$ much larger than 25. 
Furthermore, all four procedures belong to the class of 
so called `embarrassingly parallel' algorithms, i.e. 
their time complexity scales as the inverse
 of the number of processors meaning that 
they are particularly well suited to benefit
 from modern computing environments. To 
enhance user experience, we implemented 
FastPCS in \cpp code wrapped in a portable 
\texttt{R} package \citep{mcs:R} 
distributed through \texttt{CRAN} (package \texttt{FastPCS}).

\section{Empirical Comparison: Simulation Study}\label{mcs:s3}

In this section we evaluate FastPCS 
numerically and contrast its performance 
to that of the SDE, FastMCD 
and FastMVE algorithms. The last three were 
computed using the \texttt{R} package 
\texttt{rrcov} with default settings (except,
 respectively, for the number of random directions
 and starting subsets which for all algorithms we
set according to Equation $\eqref{mcs:Ns}$). 
Each algorithm returns a subset $H_*$ of $h$ 
observations with the smallest outlyingness index.
 Our evaluation criteria are the bias and the 
misclassification rate of these $h$ observations.
 Below, we briefly describe these.

\subsection{Asymptotic Bias}

Given a central model 
$\mathcal{F}_u$
 and an arbitrary contaminating distribution 
$\mathcal{F}_c$, consider 
the $\varepsilon$-contaminated model
\begin{eqnarray}
\mathcal{F}_{\varepsilon}=(1-\varepsilon)\mathcal{F}_u(\mu_u,\varSigma_u)+\varepsilon\mathcal{F}_c(\mu_c,\varSigma_c)\;.
\end{eqnarray}
For a subset of observations 
$H_*$, the (asymptotic) bias measures
 the deviation of $(t_*,S_*)$ 
from the true $(\mu_u,\varSigma_u)$. 
Formally, denoting $G=|S_*|^{-1/p}S_*$
 and $\Gamma=|\varSigma_u|^{-1/p}\varSigma_u$, 
we have that for an affine equivariant
 scatter matrix $S$, all the information about
 the bias is contained in the matrix 
$G^{-1/2} \Gamma G^{-1/2}$ or equivalently
 its condition number \citep{mcs:YM90}:
\begin{equation}\label{sbias}
\mbox{bias}(S) =
      \log\lambda_1(G^{-1/2} \Gamma G^{-1/2})-\log\lambda_p(G^{-1/2} \Gamma G^{-1/2})\;, 
\end{equation}
where $\lambda_1$ ($\lambda_p$) are the largest 
(smallest) eigenvalues of $G^{-1/2} \Gamma G^{-1/2}$. 
Evaluating the bias of $(t_*,S_*)$ is an empirical
 matter. For a given sample, the bias will depend 
on the dimensionality of the data, the rate of 
contamination and the distance separating the 
outliers from the good part of the data. 
Finally, the bias also depends on the spatial 
configuration of the outliers (the choice of 
$\mathcal{F}_c$). Fortunately, for affine equivariant
 algorithms the outlier configurations causing the 
largest biases are known and so we can focus on these
 cases. 

\subsection{Misclassification rate}

We can also compare the outlyingness 
indexes in terms of rate of contamination 
of their final $h$-subsets. Denoting by $I_c$
 the index
 set of the contaminated observations and 
$\mathcal{I}(\bullet)$ the indicator function, 
the misclassification rate is:
\begin{eqnarray}
\mbox{Mis.Rate}(I_c,H)=\frac{\sum_{i\in H}\mathcal{I}(i\in I_c)}{\sum_{i=1}^n\mathcal{I}(i\in I_c)}\;.
\end{eqnarray}
This measure is always in $[0,1]$, thus
 yielding results that are easier to 
compare across dimensions and rates of
 contamination. A value of 1 means that 
the $H$ contains all the outliers. 
The main difference with the bias 
criterion is that the misclassification
 rate does not account for how disruptive
 the outliers are. For example, when the distance separating 
the outliers from the good part of the data is small,  
it is possible for $\mbox{Mis.Rate}(I_c,H_*)$
 to be large without a commensurate 
increase in $\mbox{bias}(S_*)$. 

\subsection{Outlier configurations}
We generate many contaminated datasets
 $X_\varepsilon$ of size $n$ with 
$X_{\varepsilon}=X_u\cup X_c$ where 
$X_u$ and $X_c$ are, respectively, 
the genuine and outlying part of 
the sample. 
The bias depends on the distance between the 
outliers and the genuine observations which 
we will measure by
\begin{equation}\label{mcs:nu}
\nu = \min_{i\in I_c} \sqrt{d^2_{MD,i}(t_u,S_u)/\chi^2_{0.99,p}} \,\, .
\end{equation}
The bias also depends on the 
spatial configuration of $X_c$. For 
affine equivariant algorithms, the worst-case
configurations (those causing the largest bias) are known. 
In increasing order of difficulty these are:

\begin{itemize}
\item Shift configuration. If we constrain the adversary to 
      (a) $|\varSigma_c|\ge|\varSigma_u|$ and (b) place $X_c$ 
      at a distance $\nu$ of $X_u$, then, to maximize the bias,
      the adversary will set $\varSigma_c=\varSigma_u$ (Theorem 1 in 
       \citep{mcs:RW96}) and set $\mu_c$ in order to satisfy (b).
      Intuitively, this makes the components of the mixture the
       least distinguishable from one another.
\item Point-mass configuration. If we omit the constraint (a) 
      above but keep (b), the adversary will place $X_c$ in
      a single point so $|\varSigma_c|=0$ (Theorem 2 in
       \citep{mcs:RW96}).
\item If we omit both constraints (a) and (b), the adversary
      may set $\mu_c=\mu_u$ and choose $\varSigma_c$ to obtain
      a large bias. The Barrow wheel contamination 
       \citep{mcs:SM09} does this.
\end{itemize}

\begin{figure}[h!]
\centering
\includegraphics[width=1\textwidth]{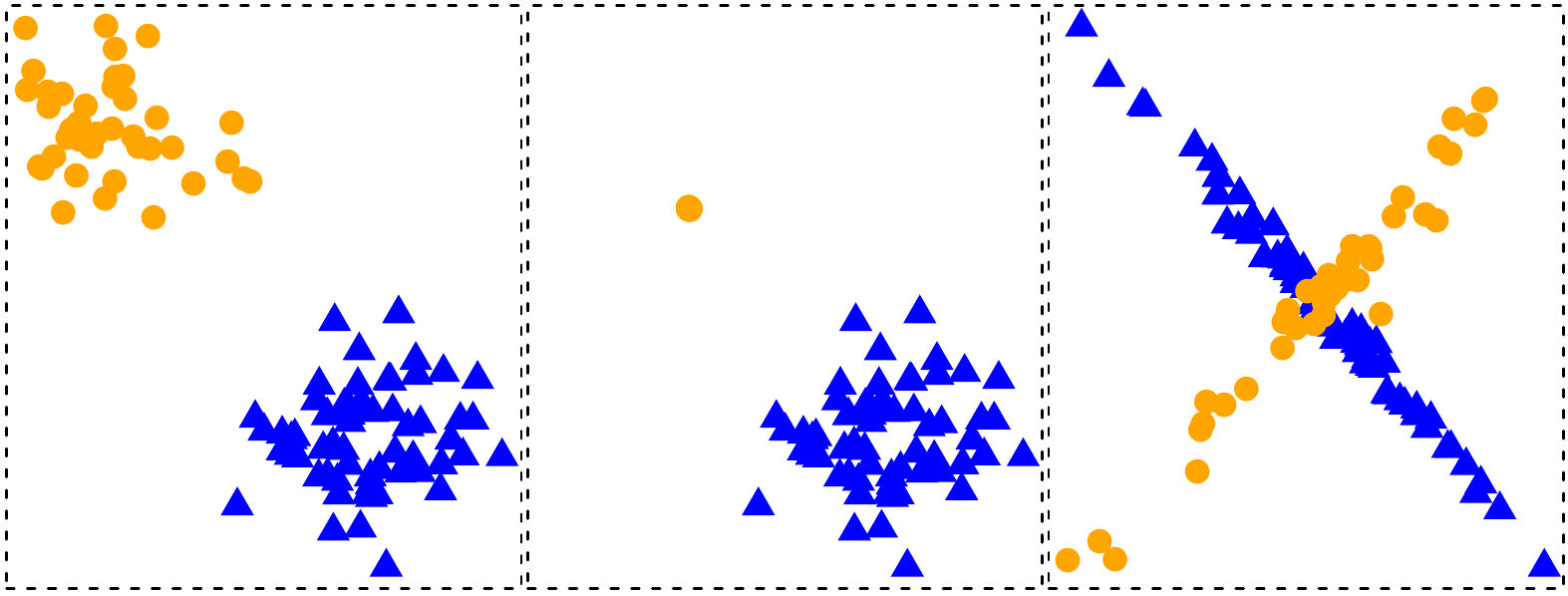}
\caption{The three outlier configurations (Shift, Point-mass and Barrow wheel).
         The outliers are depicted as light orange dots.} 	
\label{mcs:f0}
\end{figure}

We also considered other 
configurations such as 
radial outliers as well 
as cases where $\nu$ was 
set to extremely large values (i.e. $\nu\geq10000$), 
but they posed little challenge 
for any of the algorithms,
 so these results are not
 shown.
\subsection{Simulation parameters}\label{mcs:s5}
We can generate both the uncontaminated
 data $X_u$ and the contaminated data 
$X_c$  from the standard normal distribution
 since all methods under consideration are 
affine equivariant. For the shift and point 
configurations, we will also generate data 
from the standard Cauchy distribution in 
order to quantify the sensitivity of each 
method to the tail index of the data.  

For the shift and point configurations we 
generated the outliers by setting $\varSigma_c$
 as either $D_p$ or $10^{-4}D_p$ ($D_p$ denotes a rank $p$ 
diagonal matrix) and set $\mu_c$ so that 
Equation $\eqref{mcs:nu}$ is satisfied. We generated 
the Barrow wheel configuration using the \texttt{robustX} 
package \citep{mcs:SM09} with default parameters. The
 three configuration types are depicted in Figure 
\ref{mcs:f0} for $n=100$, $p=2$, $\varepsilon=0.4$, 
$\nu=2$ and $\mathcal{F}_u$ is the bivariate normal.
 The outlying observations are the lighter orange circles.  
The complete list of simulation parameters follows:

\begin{itemize}
\item the dimension $p$ is one of $\{4,8,12,16\}$,
\item the sample size is $n=25p$, 
\item the contamination fraction $\varepsilon$ is one
      of $\{0.1,0.2,0.3,0.4\}$,
\item the configuration of the outliers is either
      shift, point, or Barrow wheel,
\item for shift and point contamination, the distance $\nu$
      comes from the uniform distribution on (0,10).
      The Barrow wheel contamination does not depend on
      $\nu$.
\item $\alpha\in(0.5,1)$ is a parameter determining the size 
of the subset of observations assumed to follow a model (with 
$h\approx\lceil\alpha n\rceil)$. 
In Section~\ref{mcs:s6a} (Section~\ref{mcs:s6b}) we consider the 
case where we set $\alpha=0.5$ ($\alpha=0.75$). 
\item the number of initial $(p+1)$-subsets $M_p$ is given by \citep{mcs:S81}
      \begin{equation}\label{mcs:Ns}
        M_p=\frac{\log(0.01)}{\log(1-(1-\varepsilon_0)^{p+1})}\;,
      \end{equation}
      with $\varepsilon_0=4(1-\alpha)/5$ so that the probability of getting at least one
      uncontaminated subset is always at least 99 percent.
\end{itemize}

In Figures \ref{mcs:f1} to \ref{pcs:pm25} we display
 the bias and the misclassification rate (left and 
right plots, respectively) for discrete combinations
 of the dimension $p$, and contamination rate 
$\varepsilon$. In all cases, we expect the outlier
 detection problem to become monotonically harder as
 we increase $p$ and $\varepsilon$, so not much 
information will be lost by considering a discrete 
grid of a few values for these parameters. For the Barrow wheel, 
 $p$ and $\varepsilon$ are the only parameters we have and so 
we can chart the results as trellises of 
 boxplots \citep{mcs:D08}.

For the shift and point contamination models, the 
configurations also depend on the distance separating
 the data from the outliers. 
 Here, the effects of $\nu$ on the bias are harder to foresee: clearly 
  nearby outliers will be harder to detect but misclassifying 
distant outliers will increase the bias more. Therefore, we 
 will test the algorithms for many values (and chart 
the results as a function) of $\nu$. For both the bias and the
 misclassification rates curves, for each algorithm, a solid colored
 line will depict the median and a dotted line (of the same 
color) the 75th percentile. Here, each panel will be based on 1000 
simulations. 

\subsection{Simulation results (a)}\label{mcs:s6a}

The first part of the simulation study covers 
the case where there is no information about the
 extent to which the data is contaminated. Then,
  we have to set the size of the subset of observations 
having positive weight to $h=\lfloor(n+p+1)/2\rfloor$, corresponding to 
the lower bound of slightly more than the majority
 of the observations.

\begin{figure}[h!]
\centering
\includegraphics[width=0.49\textwidth]{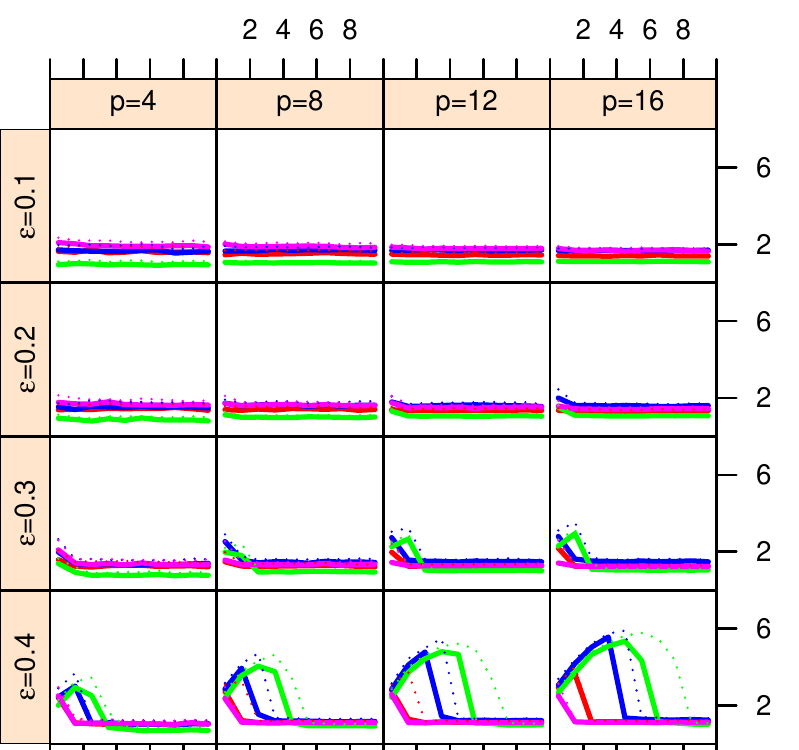}
\includegraphics[width=0.49\textwidth]{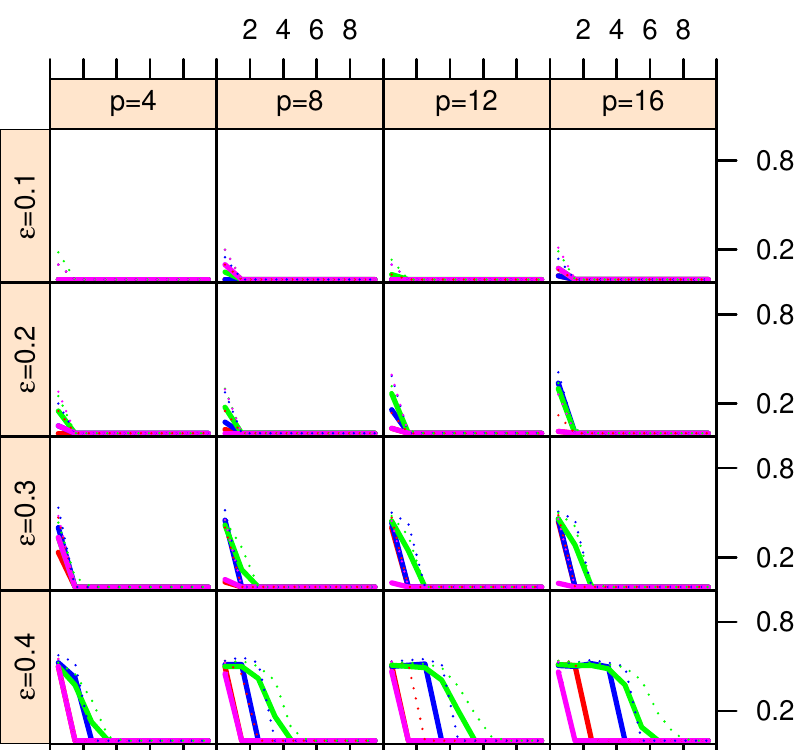}
\caption{Bias (left) and misclassification rate (right) due to shift contamination for 
         $\varepsilon=\{0.1,\ldots,0.4\}$, $p=\{4,\ldots,16\}$, $h=\lfloor(n+p+1)/2\rfloor$ and Normal-distributed 
	observations shown as a function of $\nu$.
  \textcolor{fmcd}{FastMCD} (dotdash lines),
         \textcolor{fmve}{FastMVE} (longdash lines), 
         \textcolor{msde}{SDE} (twodash lines),
         \textcolor{PCS}{FastPCS} (solid lines).}	
\label{mcs:f1}
\end{figure}

In Figure \ref{mcs:f1} 
we display the bias ($\mbox{bias}(S_*)$) 
and the misclassification rate 
($\mbox{Mis.Rate}(I_c,H_*)$) curves 
of each algorithm as a function of $\nu$ for
 different values of $p$ and $\varepsilon$  for the
 shift configuration when $\mathcal{F}_u$ is the 
standard normal. All the methods perform well until
 $\varepsilon=0.3$ and $p\leq8$. For larger values
 of $p$ and $\nu$, the bias curves and misclassification
 rate of FastMVE and SDE clearly stand out for values of
 $\nu$ between 2 and 6. Eventually, as we move the outliers
 further away, SDE and FastMVE can again find them 
reliably.
 
\begin{figure}[h!]
\centering
\includegraphics[width=0.49\textwidth]{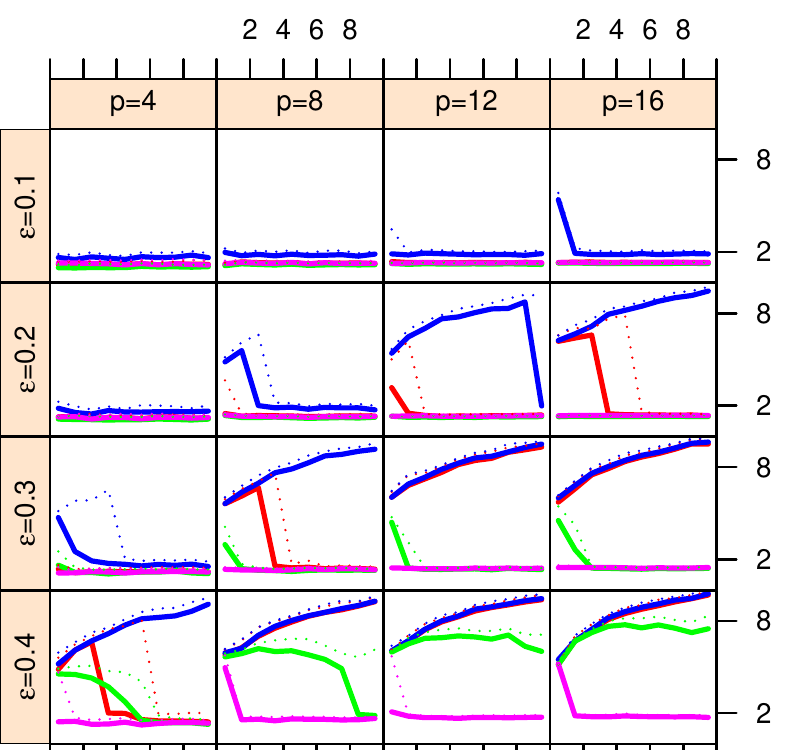}
\includegraphics[width=0.49\textwidth]{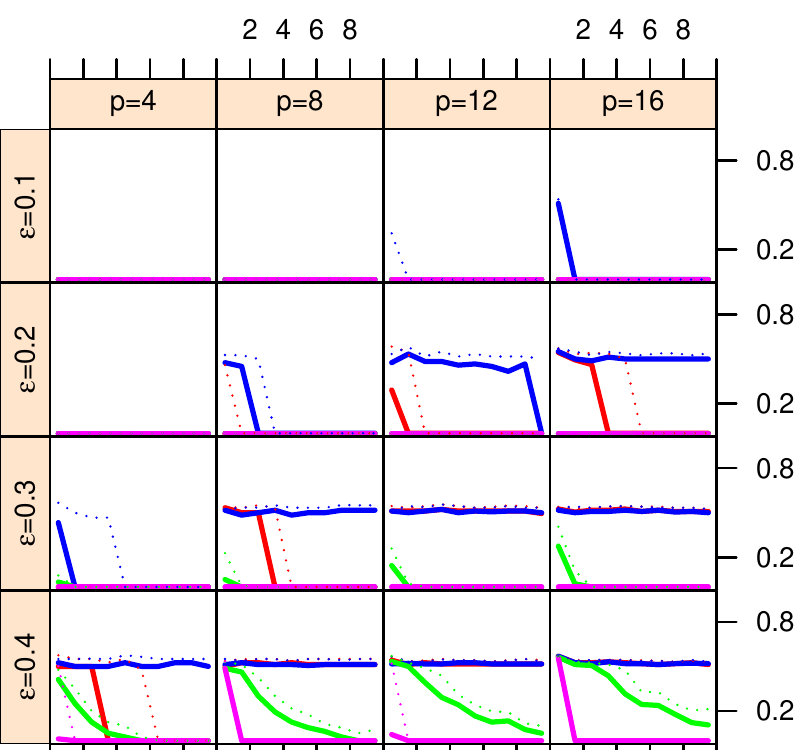}
\caption{Bias (left) and misclassification rate (right) due to shift contamination for 
         $\varepsilon=\{0.1,\ldots,0.4\}$, $p=\{4,\ldots,16\}$, $h=\lfloor(n+p+1)/2\rfloor$ and Cauchy-distributed 
	observations shown as a function of $\nu$.
  \textcolor{fmcd}{FastMCD} (dotdash lines),
         \textcolor{fmve}{FastMVE} (longdash lines), 
         \textcolor{msde}{SDE} (twodash lines),
         \textcolor{PCS}{FastPCS} (solid lines).}	
\label{mcs:f2}
\end{figure}

When $\mathcal{F}_u$ is Cauchy (Figure \ref{mcs:f2}), 
FastMVE performs significantly worse than the other 
algorithms starting already at $\varepsilon=0.2$ and 
$p\ge12$. Eventually, ($\varepsilon=0.3$ and 
$p\ge12$) FastMCD also breaks away from $\varSigma_u$. 
SDE performs better until $\varepsilon=0.4$, where it is 
also noticeably affected by the thicker tails of the 
Cauchy distribution and fails to identify the outliers. 
FastPCS maintains constant and low bias (and misclassification rates) 
throughout.

\begin{figure}[h!]
\centering
\includegraphics[width=0.49\textwidth]{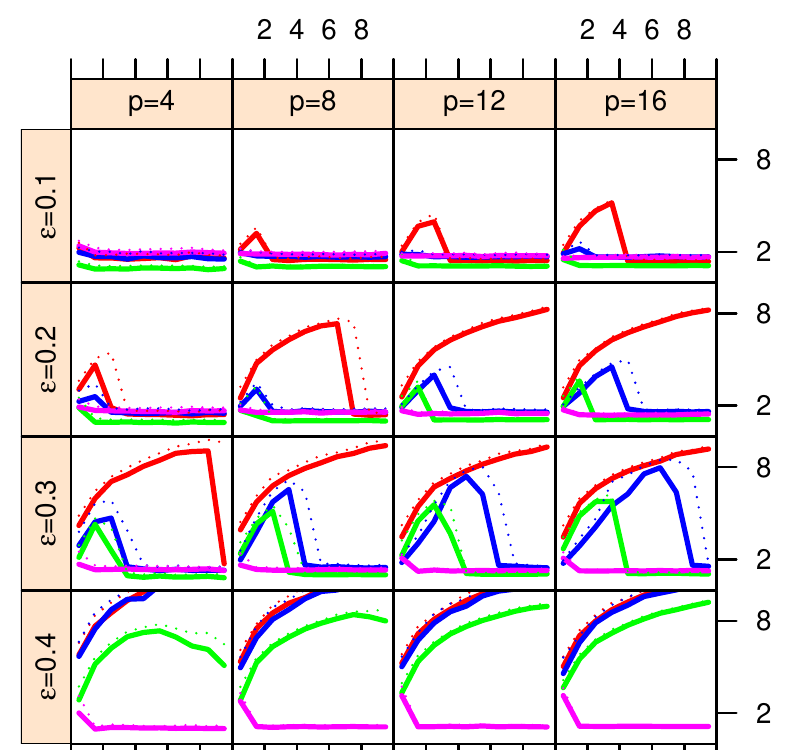}
\includegraphics[width=0.49\textwidth]{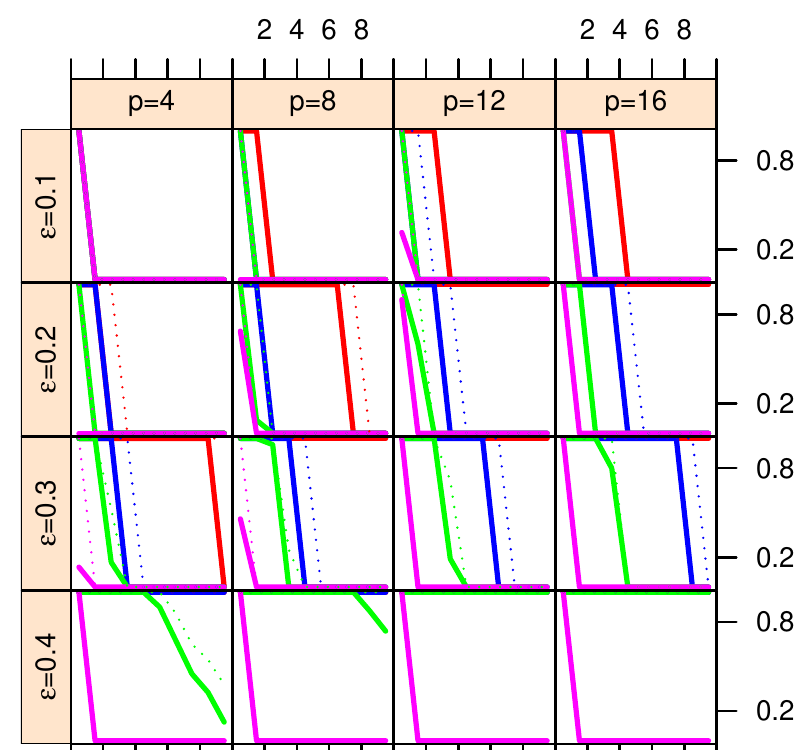}
\caption{Bias (left) and misclassification rate (right) due to point-mass contamination for 
         $\varepsilon=\{0.1,\ldots,0.4\}$, $p=\{4,\ldots,16\}$, $h=\lfloor(n+p+1)/2\rfloor$ and Normal-distributed 
	observations shown as a function of $\nu$.
  \textcolor{fmcd}{FastMCD} (dotdash lines),
         \textcolor{fmve}{FastMVE} (longdash lines), 
         \textcolor{msde}{SDE} (twodash lines),
         \textcolor{PCS}{FastPCS} (solid lines).}	
\label{mcs:f3}
\end{figure}

In Figure \ref{mcs:f3}, we show the results for the more difficult case
 of point-mass contamination when $\mathcal{F}_u$ is the
 standard normal. Starting at $\varepsilon=0.2$, the bias
 curves and misclassification rate of FastMCD become very
 high (except for $p=4$). Starting at ($\varepsilon=0.3$ 
and $p\ge8$) all the algorithms except FastPCS fail 
to reliably find the outliers: looking at the misclassification
 rate, the optimal $h$-subsets for SDE, FastMVE and FastMCD even
 contain a higher contamination rate than $\varepsilon$. 

\begin{figure}[h!]
\centering
\includegraphics[width=0.49\textwidth]{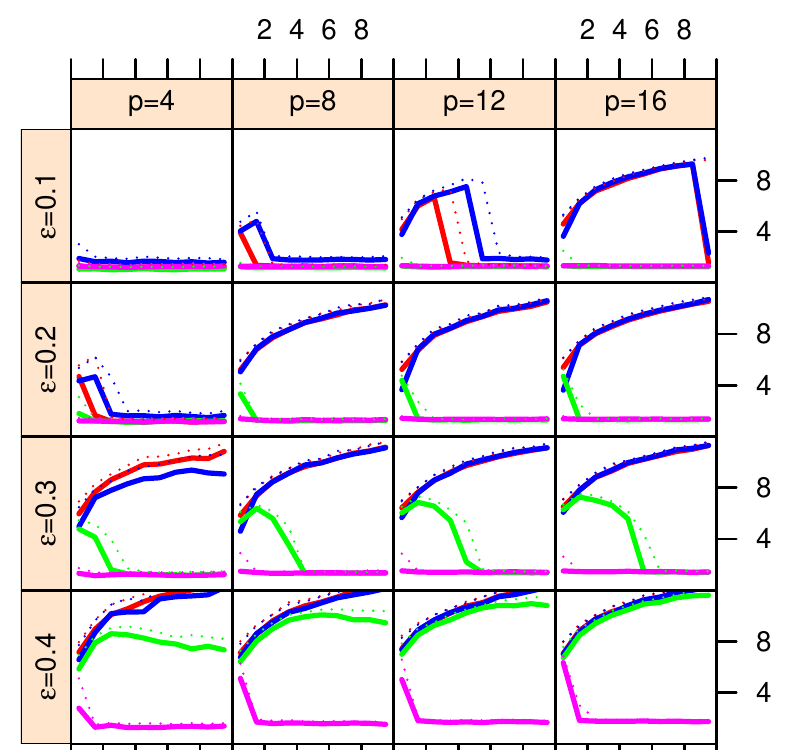}
\includegraphics[width=0.49\textwidth]{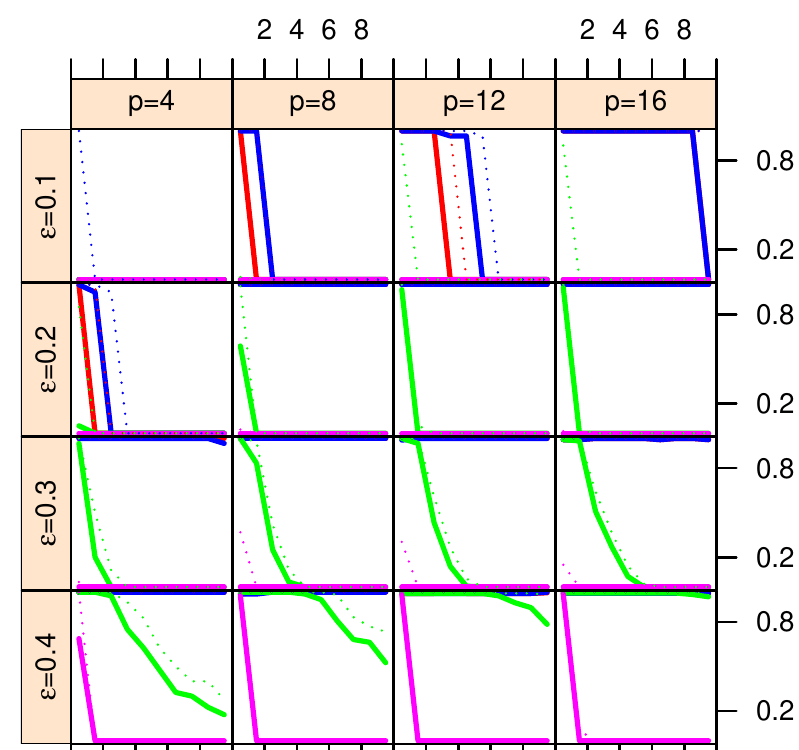}
\caption{Bias (left) and misclassification rate (right) due to point-mass contamination for 
         $\varepsilon=\{0.1,\ldots,0.4\}$, $p=\{4,\ldots,16\}$, $h=\lfloor(n+p+1)/2\rfloor$
	 and Cauchy-distributed 
	observations shown as a function of $\nu$.
        \textcolor{fmcd}{FastMCD} (dotdash lines),
         \textcolor{fmve}{FastMVE} (longdash lines), 
         \textcolor{msde}{SDE} (twodash lines),
         \textcolor{PCS}{FastPCS} (solid lines).}	
\label{mcs:f4}
\end{figure}

The Cauchy case shown in Figure \ref{mcs:f4} is also interesting. It 
suggests, again, that FastMCD and FastMVE are very sensitive to
 fat tails in the distribution of the good part of the data. The
 behavior of SDE is roughly in line with Figure \ref{mcs:f3}. 
Again, we see that FastPCS is the only algorithm that can reliably
 find the outliers. Furthermore, the misclassification rates 
reveal that the $H_*$ found by SDE, FastMCD and FastMVE often 
contains a proportion of outliers higher than $\varepsilon$.

\begin{figure}[h!]
\centering
\includegraphics[width=0.49\textwidth]{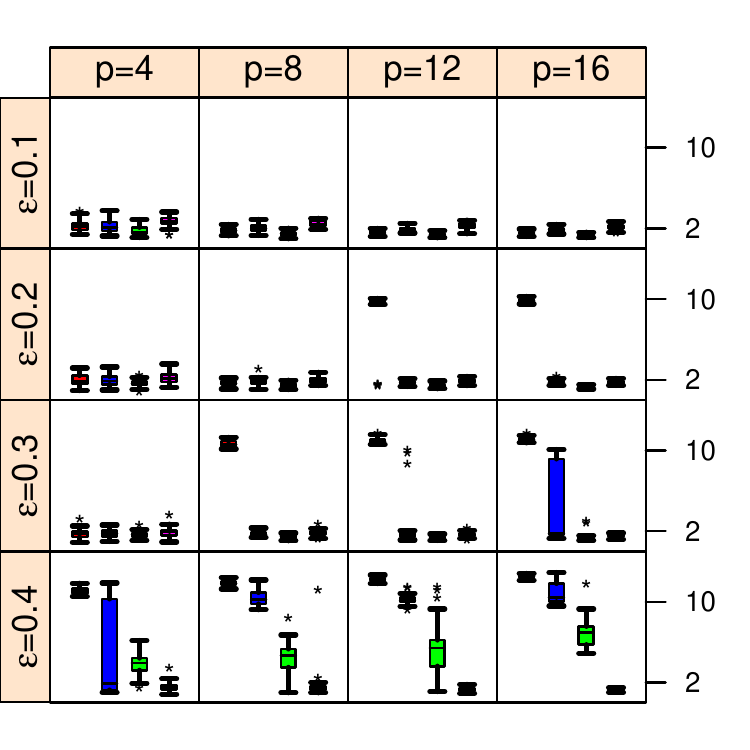}
\includegraphics[width=0.49\textwidth]{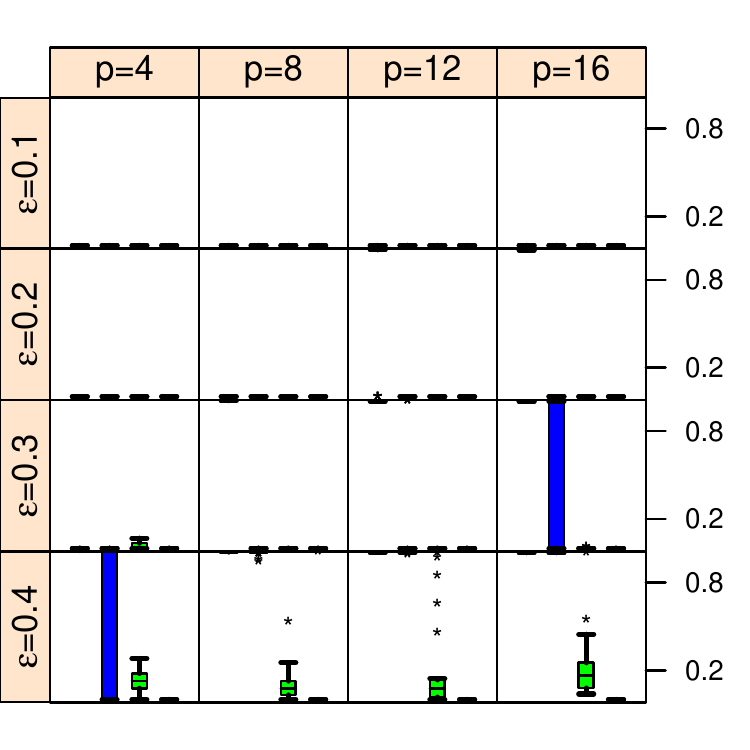}
\caption{Bias (left) and misclassification rate (right) due to the Barrow wheel contamination for 
         $\varepsilon=\{0.1,\ldots,0.4\}$ and $p=\{4,\ldots,16\}$, $h=\lfloor(n+p+1)/2\rfloor$.
         \textcolor{fmcd}{FastMCD},
        \textcolor{fmve}{FastMVE}, 
        \textcolor{msde}{SDE},
        \textcolor{PCS}{FastPCS} (far right).}	
\label{mcs:f5}
\end{figure}

In Figure \ref{mcs:f5} we show the bias curves
 for the Barrow wheel as boxplots. 
FastMCD performs badly compared to the
 other methods, and FastMVE deteriorates
 for $\varepsilon$ larger than 30 percent.
 It is at $\varepsilon=0.4$ that FastPCS
 demonstrably exhibits superior performance,
 with SDE producing results in between 
the other three algorithms.

Across the tests, FastPCS consistently maintains low
 and stable biases and misclassification rates. 
Furthermore, the performance of FastPCS is relatively
 insensitive to whether $\nu$, $\varepsilon$ and $p$
  are low or high. Additionally, in contrast with the 
other algorithms, FastPCS is unaffected by the thickness
 of the tails of $\mathcal{F}_u$. The consistent ability 
of FastPCS to detect outliers is a highly desirable feature, 
which we characterize as a form of qualitative robustness.

\subsection{Simulation results (b)}\label{mcs:s6b}
In this section, we consider the case, important 
in practice, where the user can confidently place 
an upper bound on the rate of contamination of the 
sample. To fix ideas, we will set $h$, the 
number of observations assumed to follow a model,  
to $h\approx3n/4$. For FastPCS, FastMVE and FastMCD
 we adapt the algorithms by setting their respective 
$\alpha$ parameter to 0.75. 
SDE does not have a corresponding parameter, so 
we take as member of $H_*$ the (approximately) $3n/4$ 
observations with smallest outlyingness. Furthermore, 
we reduce the number of starting subsets (FastPCS, FastMCD, 
FastMVE) and random directions (SDE), by using 
Equation $\eqref{mcs:Ns}$, but this time setting $\varepsilon_0=0.2$.
 Then, as before, we measure the effect of our various configurations of
 outliers on the estimators (but this time only considering 
 values of $\varepsilon\in\{0.1,0.2\}$). 

\begin{figure}[h!]
\centering
\includegraphics[width=0.49\textwidth]{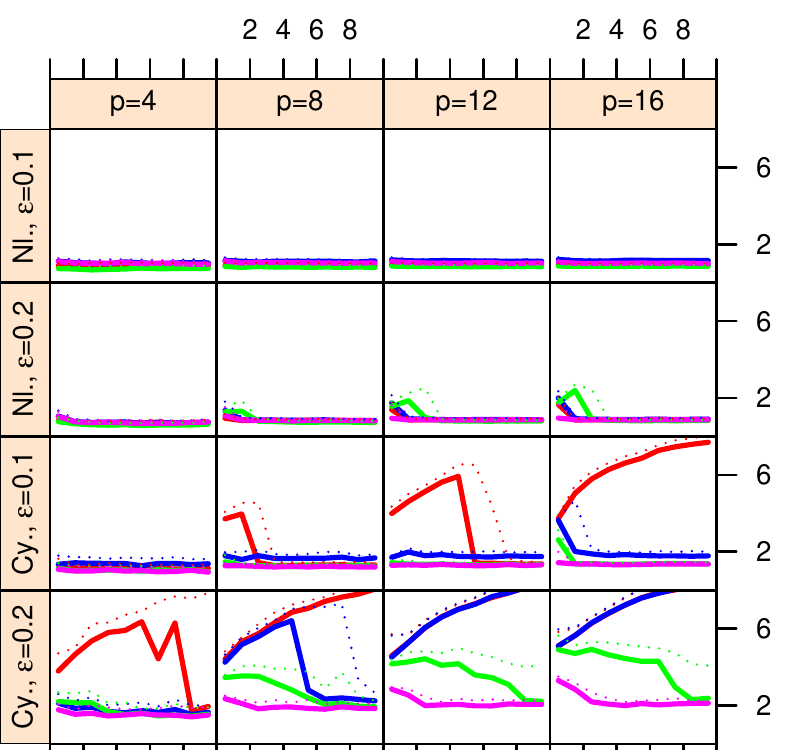}
\includegraphics[width=0.49\textwidth]{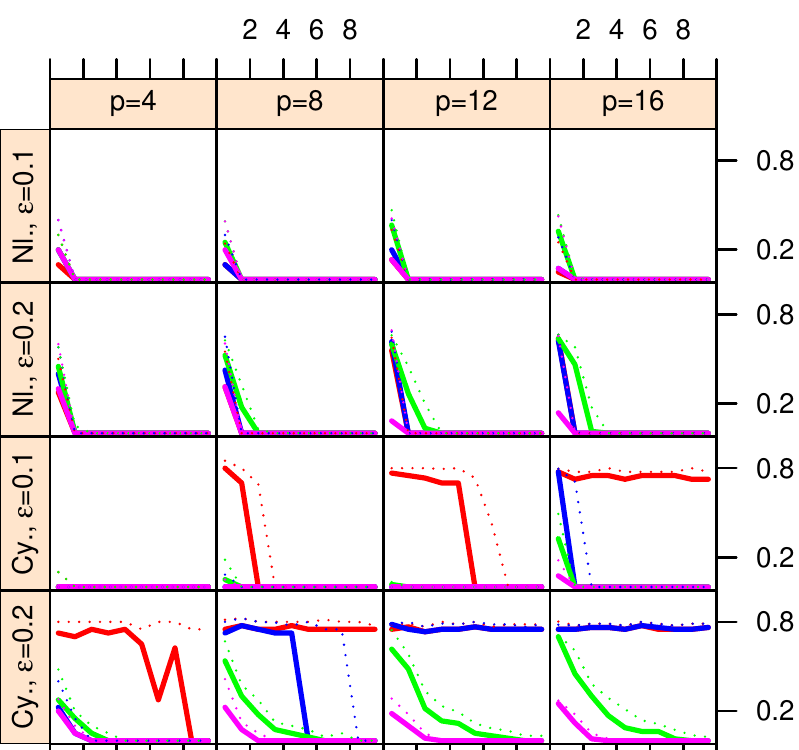}

\caption{Bias (left) and misclassification rate (right) due to the shift contamination for 
         $h\approx 3n/4$, $\varepsilon=\{0.1,0.2\}$ and $p=\{4,\ldots,16\}$. 
	The first (last) two rows are for multivariate normal (Cauchy) distributed r.v. 
         \textcolor{fmcd}{FastMCD} (dotdash lines),
         \textcolor{fmve}{FastMVE} (longdash lines), 
         \textcolor{msde}{SDE} (twodash lines),
         \textcolor{PCS}{FastPCS} (solid lines).}	
\label{pcs:shift25}
\end{figure}

\begin{figure}[h!]
\centering
\includegraphics[width=0.49\textwidth]{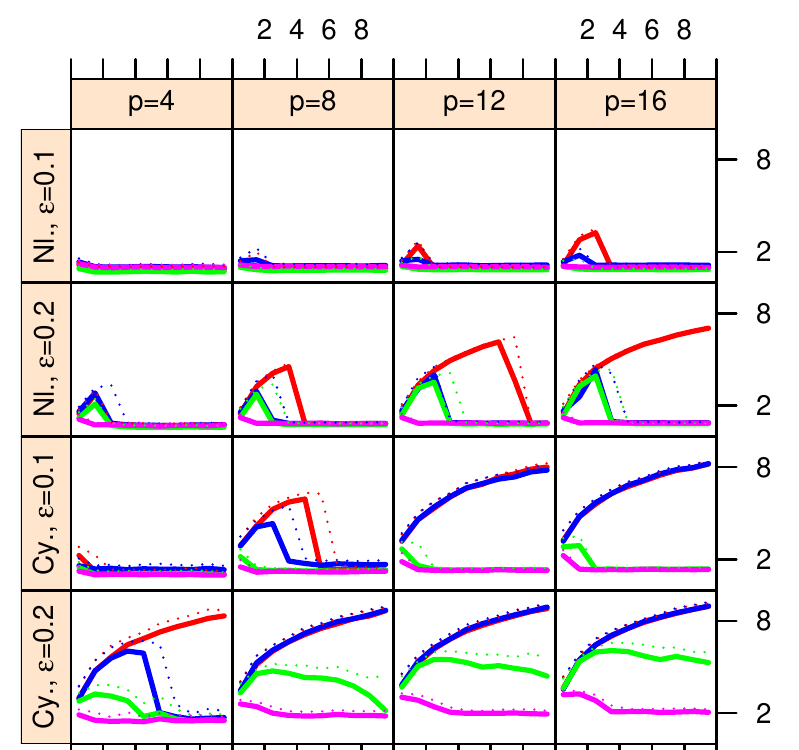}
\includegraphics[width=0.49\textwidth]{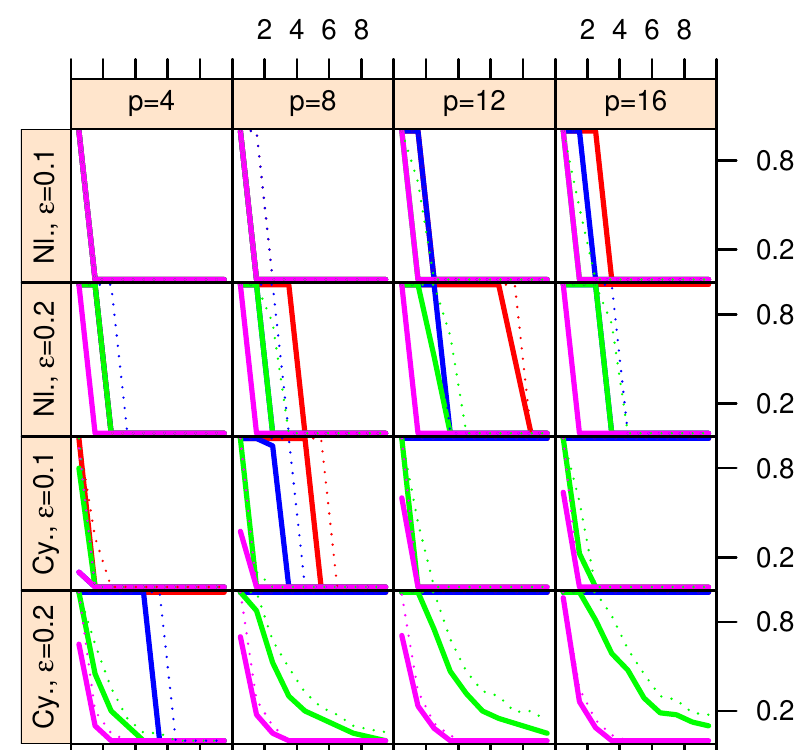}

\caption{Bias (left) and misclassification rate (right) due to point-mass contamination for 
         $h\approx 3n/4$, $\varepsilon=\{0.1,0.2\}$ and $p=\{4,\ldots,16\}$.
	The first (last) two rows are for multivariate normal (Cauchy) distributed r.v. 
         \textcolor{fmcd}{FastMCD} (dotdash lines),
         \textcolor{fmve}{FastMVE} (longdash lines), 
         \textcolor{msde}{SDE} (twodash lines),
         \textcolor{PCS}{FastPCS} (solid lines). The online version of this figure is in color.}	
\label{pcs:pm25}
\end{figure}

In Figures~\ref{pcs:shift25} and ~\ref{pcs:pm25}, we show the simulation results for shift and point-mass contamination. The results for Normal-distributed observations are shown in the first two rows, while the 
last two rows show the results for observations drawn from the multivariate Cauchy distribution. 
The shift contamination results in Figure~\ref{pcs:shift25} show that when we increase the size of the 
active subsets, FastPCS still maintains a low bias and misclassification rate, but at equivalent rates 
of contamination, the other algorithms exhibit noticeably weaker performance than in Figures~\ref{mcs:f1} 
and \ref{mcs:f2}. Results in Figure~\ref{pcs:pm25} illustrate that under point-mass contamination, FastPCS again reports good results. The other algorithms continue to exhibit larger biases and misclassification rates (at equivalent rates of contamination) than in Figures \ref{mcs:f3} and \ref{mcs:f4}. We do not show the results for the Barrow wheel because all the algorithms perform equivalently.

\subsection{Small Sample Accuracy}

For affine equivariant outlier detection algorithms
 there is a trade-off between bias and
 accuracy \citep{mcs:R94}. This is especially
 problematic in small $n$ and low $\varepsilon$
 settings because then any reduction in bias can
 be partially undone by losses in small sample
 accuracy. For FastPCS in particular, this is 
visible in the first rows of Figures \ref{mcs:f1}
 and \ref{mcs:f5}: in the misclassification plots,
 we see that for all algorithms the optimal subsets
 $H_*$ are uncontaminated by outliers. Yet, the 
corresponding bias curves for FastPCS tend to be 
somewhat higher than for the other algorithms. 

To improve accuracy in small samples 
without conceding too much in bias, a solution 
 is to add a so-called re-weighting step 
to the outlier detection algorithm. 
In essence, we replace $H_*$ by a larger 
subset $J_+$, itself derived from 
$H_*$. The motivation is that, typically, 
 $J_+$ will include a greater 
 share of the uncontaminated observations.
 Over the years, many such
re-weighting procedures have been proposed \citep{mcs:MMY06}. 
Perhaps the simplest is the so called
 hard-thresholding re-weighting. 
Given an optimal subset $H_*$,
 the members of $J_+$ are: 
\begin{eqnarray}\label{msc:hp}
\;\;\;\;\;J_+=\left\{i:d_{MD,i}^2(t_*,S_*)\le\chi^2_{0.975,p}\left(\med_{i=1}^{n}d_{MD,i}^2(t_*,S_*)/\chi^2_{0.5,p}\right)\right\}\;,
\end{eqnarray}
where the ratio on the rhs of the inequality  
is a scale factor we use to make $S_*$
 consistent at the normal distribution. 

\begin{figure}[h!]
\centering
\includegraphics[width=0.98\textwidth]{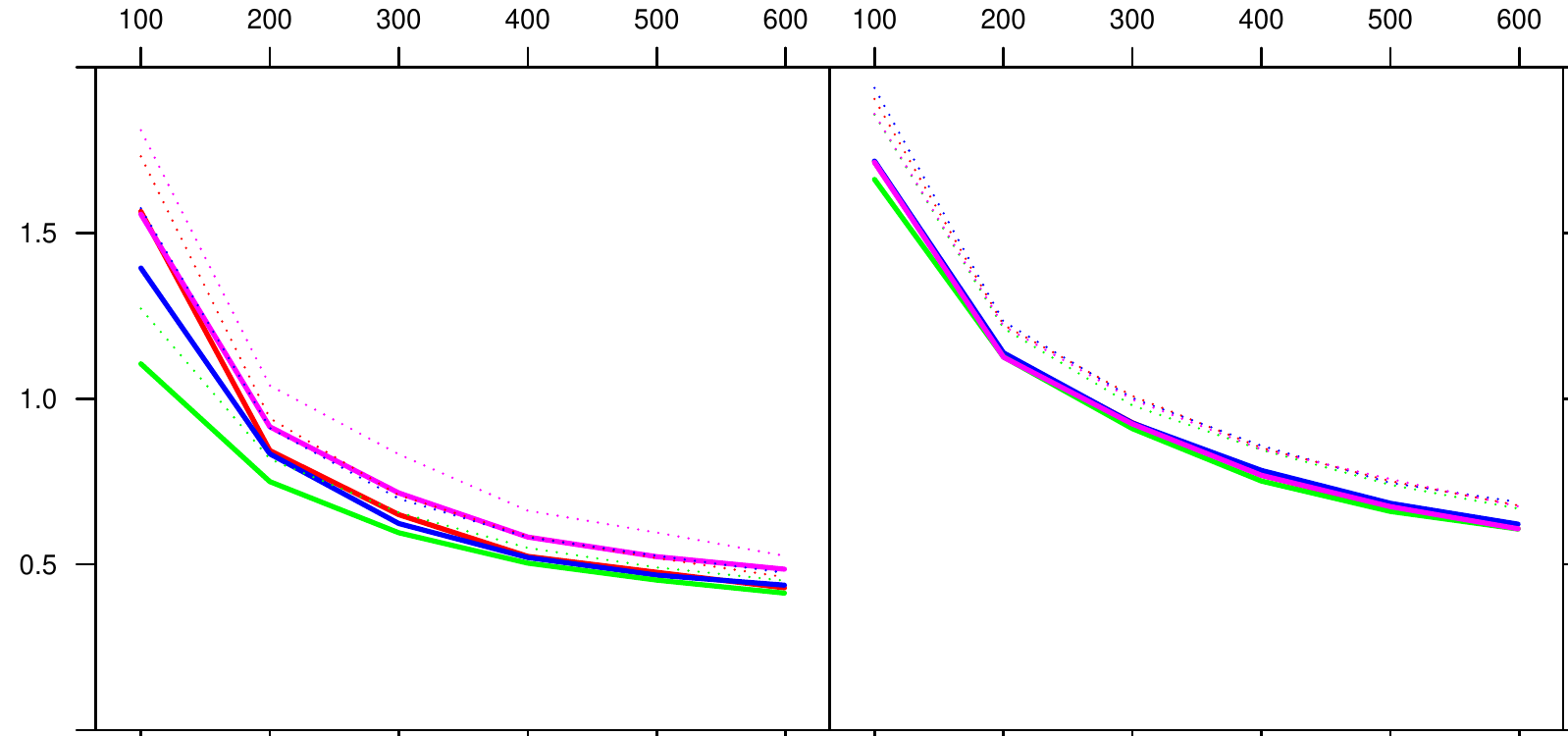}
\caption{Empirical bias at uncontaminated samples for $p=8$ as a function of sample size. 
         \textcolor{fmcd}{FastMCD} (dotdash lines),
         \textcolor{fmve}{FastMVE} (longdash lines), 
         \textcolor{msde}{SDE} (twodash lines),
         \textcolor{PCS}{FastPCS} (solid lines). The online version of this figure is in color.}
\label{mcs:f6}
\end{figure}

Contrary to $H_*$, the size of $J_+$
 is not fixed in advance. This makes
 it difficult to compare algorithms 
on the basis of $\mbox{bias}(S_+)$ 
or $\mbox{Mis.Rate}(I_c,J_+)$ evaluated
 at contaminated sub-samples $X_{\varepsilon}$.
 Nonetheless, we can still compare 
them in terms of the  bias of $S_+$ 
computed at uncontaminated data sets
 $X_u$ and for various sample sizes
 $n$. In essence, we measure the 
 accuracy of the 
various algorithms by considering 
their biases at uncontaminated 
datasets as a function of $n$. 
In Figure \ref{mcs:f6}, we show
 the results of doing this for 
$n=\{100,200,...,500,599\}$ 
(shown as the abscissa) and $p=8$. 
We set $599$ observations as an upper limit  
because when $n\ge600$ FastMCD  
 uses a nested sub-sampling scheme whereby 
 larger datasets are divided unto non-overlapping sub-samples of at most 600 
observations. 

Figure \ref{mcs:f6} depicts the median (solid)
 and 75th percentile (dotted) accuracy curves
 of the algorithms for increasing $n$ when 
$\mathcal{F}_u$ is normally distributed (left) 
and Cauchy distributed (right). Each panel is 
based on 1,000 experiments. As expected, FastPCS is noticeably
 less accurate than the other algorithms in the 
normal case. Nonetheless, we maintain that 
this difference is small compared to the reduction
 in bias achieved by FastPCS on contaminated samples.
 Furthermore, this gap in performance depends on the
 distribution of the good data: when $\mathcal{F}_u$
 is Cauchy the accuracy of FastPCS is similar to that
 of the other algorithms.

\section{Empirical Comparison: Case Study}\label{mcs:s4}

In this section, we illustrate the behavior
 of FastPCS on a real data problem from the
 field of engineering: the Concrete Slump 
Test Data set \citep{mcs:Y07}. This dataset
 includes 103 data-points, each 
corresponding to a different type of concrete. 
 For each observation we have 7 so-called
 input variables measuring,
 respectively, the quantity of 
cement (kg/m3), fly ash (kg/m3),
 blast furnace slag (kg/m3), 
water (kg/m3), super-plasticizer
 (kg/m3), coarse aggregate (kg/m3),
 and fine aggregate (kg/m3) used
 to make the corresponding variety
 of concrete. Additionally, for 
each observation, we have
 3 so-called output variables 
measuring attributes of the 
resulting concrete. These are 
slump (cm), flow (cm) and 28-day
 compressive strength (Mpa). 

This dataset actually contains two groups
 of observations collected over separate 
periods. The first 78 measurements pre-date
 the last 25 by several years. An interesting
 aspect of this dataset is that these two 
groups largely overlap on bivariate scatter-plots
 of the data, forming a seemingly homogeneous 
group of observations.
 Appearances can be deceptive however: when considered
 along all variables jointly, the two groups
 are clearly distinct. 
For example, denoting $J_O$ ($J_N$) the subset
 of 78 (25) members of the early (latter) batch
 of measurements, we find that the closest member
 of $J_N$ lies at a squared Mahalanobis distance of over 
760 with respect to $(t_O,S_O)$. As a point of 
comparison, this is nearly 32 times larger than 
$\chi^2_{0.99,10}$. 

We first ran the SDE, FastMCD, FastMVE and FastPCS
 algorithms on the raw dataset (for the first three,
 we used the \texttt{rrcov} implementations 
with default parameters). We set the number of 
random directions in SDE and the number of
 random $p$-subsets in FastMCD, FastMVE 
and FastPCS to $M_p=2000$, as given in Equation $\eqref{mcs:Ns}$.
To have comparable results, we
will use for each estimator $d_{MD,i}(t_*,S_*)$,
the vector of statistical distances wrt $H_*$.

\begin{figure}[h!]
\centering
\includegraphics[width=1\textwidth]{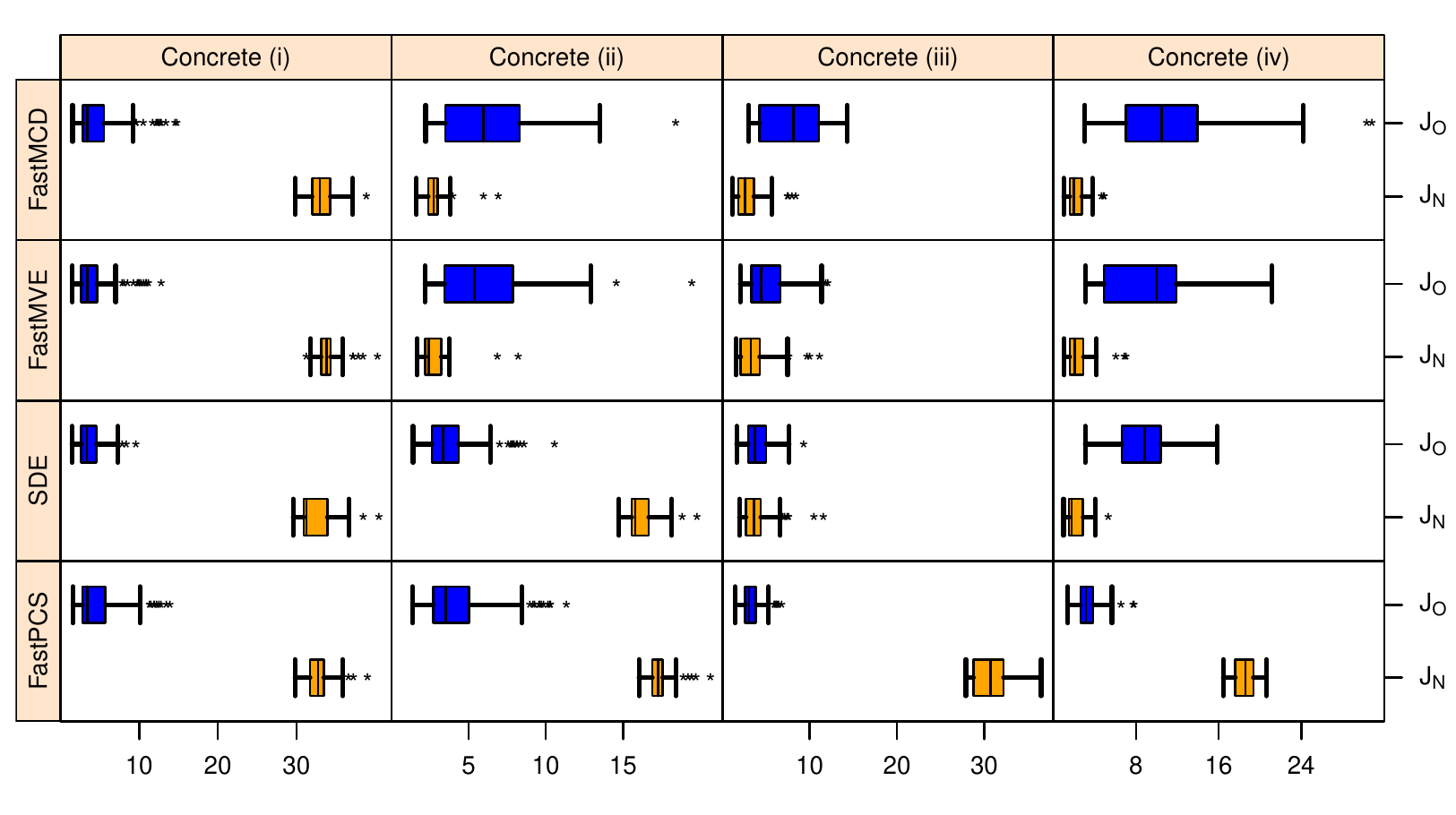}
\caption{Mahalanobis outlyingness index $d_{MD,i}(t_*,S_*)$ for the four estimators and the 
four variants of the concrete dataset. In each panel, the dark blue (light orange) boxplot depicts 
$d_{MD,i}(t_*,S_*)$ for the members of $J_O$ ($J_N$). The online version of this figure is in color.}
\label{mcs:f7}
\end{figure}

The Mahalanobis outlyingness indexes are displayed 
 in the first column of Figure \ref{mcs:f7},
 Concrete (i). Each row corresponds to an estimator.
 For each estimator we drew two boxplots. 
The first one (dark blue) depicts the outlyingness index 
for the 78 members of $J_O$ and the second (light orange) 
boxplot for those 25 members of $J_N$. All the 
algorithms are able to unambiguously separate
 the two subgroups of observations: the 
outlyingness values assigned to the members of 
the more recent batch of observations are notably
 larger than the outlyingness values assigned
 to any members of $J_O$. 

In a second experiment, Concrete(ii), we made the original
 outlier detection problem harder by narrowing the distance
 separating the outliers from the genuine observations. 
To do so, we pulled the 25 outliers towards the mean of 
the good part of the data by replacing the original outliers
 with 25 observations of the form $x_{n+i}^{ii}\leftarrow(x_{n+i}+t_O)/2$ for 
$1\le i\le 25$, $n=78$. We denote $J_N^{ii}$ to 
be these 25 members of $x_{n+i}$ for $1\le i\le 25$. 
The second column of Figure \ref{mcs:f7} depicts the outlyingness 
indexes for this second experiment. Again, for each algorithm, 
the first (dark blue) boxplots depict the values of the outlyingness 
indexes for the 78 members of $J_O$ and the second (lighter, orange) boxplot
 for the 25 members of $J_N^{ii}$. Note that the members of 
$J_N^{ii}$ still form a cluster that is genuinely separate from the 
main group of observations. For example, the closest member of
 $J_N^{ii}$ lies at a squared Mahalanobis distance of over 190 with 
respect to $(t_O,S_O)$. For comparison, this is over 8 times 
larger than $\chi^2_{0.99,10}$. Here, the FastPCS and SDE 
outlyingness indexes continue to clearly distinguish between
 the two groups. The outlyingness indexes derived from  
FastMVE and FastMCD however fail to adequately flag the outliers.
  
In a third experiment, Concrete(iii), we made the 
original outlier detection problem harder still by increasing 
the contamination rate of the sample. This was done by 
adding an additional 25 outliers of the form 
$x_{n+25+j}^{iii}\leftarrow(x_{n+1}+x_{n+1+j})/2$ for $1\le j\le 25$,
 $n=78$. We denote $J_N^{iii}$ the members of this third set of
 50 outliers. The third column of Figure \ref{mcs:f7} depicts 
the outlyingness indexes for this experiment. Again for each 
algorithm, a (dark) blue boxplot pertains to the 78 members of $J_O$
 and an lighter orange one to the 50 members of $J_N^{iii}$. For SDE,
 FastMVE and FastMCD we can see that increasing the contamination
 rate of the sample causes the outlyingness indexes of the two
 groups to overlap while the outlyingness index produced by 
FastPCS continues to make a clear distinction. Remarkably, the
 subsets of $h$ observations with the smallest outlyingness 
chosen by the first three algorithms are primarily composed of outliers.

In a final experiment, Concrete(iv), we narrowed the distance between
 the good data and the outliers (as in Concrete(ii)), and 
increased the contamination rate (as in Concrete(iii)). The 
fourth column of Figure \ref{mcs:f7} depicts the outlyingness 
indexes for this experiment, again with for each
algorithm a (darker) blue boxplot for the 78 members of $J_O$ and an 
lighter, orange one for the 50 members of $J_N^{iv}$. Here again, the 
 outlyingness index of SDE, FastMCD and FastMVE 
 fails to adequately flag the outliers. However, as with each 
of the previous experiments, FastPCS still assigns a larger 
index of outlyingness  to the members of the outlying group.

Overall, we see that the results of the real data experiment
 confirm those of the simulations, at least qualitatively. When the
 contamination rates are small and the outliers well separated
 from the good part of the data, all outlier detection methods
 seem to perform equally well. As we consider more difficult 
outlier settings however, we see that the FastPCS outlyingness
 index is the only one that identifies the outliers
 reliably.  

\section{Outlook}
   
In this article we introduced PCS, 
a new outlyingness index and FastPCS,
 a fast and affine equivariant algorithm 
for computing it. Like many other outlier
 detection algorithms, the performance   
of FastPCS hinges crucially on correctly
 identifying an $h$-subset of uncontaminated
 observations. Our main contribution is to
 characterize this $h$-subset using a new
  measure of the congruence of a multivariate 
  cloud of points. The main feature of this new 
characterization is that is was 
designed to be insensitive to the 
configuration of the outliers.

In our simulations, we have 
focused on configurations of outliers that 
are worst-case for affine-equivariant 
algorithms, and found that FastPCS behaves notably
 better than the other procedures we 
considered, often revealing outliers 
that would not have been identified by 
the other approaches. In practice, admittedly, 
contamination patterns will not always
 be as difficult as those we considered above 
and in many cases the different methods 
will, hopefully, concur. 
Nevertheless, given that in practice we do not
 know the configuration of the outliers, as
 data analysts, we prefer to carry our inferences
 while planing for the worst contingencies. 

Also through simulations, we 
found that the performance of
FastPCS is not affected much 
by the tail index of the
 majority of the data. For example,
 FastPCS is capable of finding
the outliers both when the
majority of the data is distributed
multivariate normal, or is drawn from
a heavy-tailed distribution, such as
 the multivariate Cauchy.

In this article we emphasized the practical
 aspects of PCS. A number of theoretical 
properties deserve further investigation. In
 particular, we suspect that PCS has maximum
 breakdown point. Arguments supporting this 
conjecture are that the maximum biases are 
low and that the procedure has the exact fit
 property.

\bibliographystyle{model2-names}







\end{document}